\documentclass[12pt,preprint]{aastex}

\received{}
\accepted{}
\slugcomment{To appear in THE ASTROPHYSICAL JOURNAL LETTERS}

\shorttitle{TIMESCALE FOR RUNAWAY MERGING OF BLACK HOLES}
\shortauthors{MOURI \& TANIGUCHI}

\begin{document}

\title{RUNAWAY MERGING OF BLACK HOLES: ANALYTICAL CONSTRAINT ON THE TIMESCALE}

\author{HIDEAKI MOURI}
\affil{Meteorological Research Institute, Nagamine 1-1, Tsukuba 305-0052, Japan; hmouri@mri-jma.go.jp}

\and

\author{YOSHIAKI TANIGUCHI}
\affil{Astronomical Institute, Graduate School of Science, Tohoku University, Aoba, Sendai 980-8578, Japan; tani@astr.tohoku.ac.jp}

\notetoeditor{You may find the expression ``$10^0$'' etc. in our order-of-magnitude discussion. Please do not replace it with, e.g., ``1''.}

\begin{abstract}

Following the discovery of a black hole (BH) with a mass of $10^3$--$10^6\,M_\sun$ in a starburst galaxy M82, we study formation of such a BH via successive merging of stellar-mass BHs within a star cluster. The merging has a runaway characteristic. This is because massive BHs sink into the cluster core and have a high number density, and because the merging probability is higher for more massive BHs. We use the Smoluchowski equation to study analytically the evolution of the BH mass distribution. Under favorable conditions, which are expected for some star clusters in starburst galaxies, the timescale of the runaway merging is at most of order 10$^7$ yr. This is short enough to account for the presence of a BH heavier than $10^3\,M_{\sun}$ in an ongoing starburst region.

\end{abstract}


\section{INTRODUCTION}

While black holes (BHs) arising from stellar evolution have masses of orders $10^0$--$10^1\,M_{\sun}$, those found in galaxy nuclei have masses of orders 10$^6$--10$^9\,M_{\sun}$ (Kormendy \& Richstone 1995; Kormendy \& Gebhardt 2001). The possible missing link, i.e., an intermediate-mass BH, was recently discovered in a starburst galaxy M82 as a source of compact X-ray emission (Kaaret et al. 2001; Matsumoto et al. 2001; see also earlier references therein). The observed strong variability implies that the source is an accreting BH. The observed luminosity of $10^{41}$ ergs s$^{-1}$ implies that the BH mass is greater than $10^3\,M_{\sun}$ if the emission is isotropic and its luminosity is below the Eddington limit. The BH lies at the 2 \micron\ secondary peak (Matsushita et al. 2000), which is an unusually active site of ongoing star formation. Hence the BH was formed during the ongoing starburst and is less massive than a so-called super star cluster, i.e., a building block of a starburst region. The typical mass of a super star cluster is of order $10^6\,M_{\sun}$.

The formation of an intermediate-mass BH is attributable to successive merging of stellar-mass BHs within a super star cluster (Taniguchi et al. 2000; Matsushita et al. 2000; Ebisuzaki et al. 2001; Miller \& Hamilton 2001). The cluster sinks into the galaxy center due to dynamical friction and upon evaporation releases the intermediate-mass BH. If there are more than one such BHs, they merge with each other. Thus an intermediate-mass BH could eventually evolve to a supermassive BH and is crucial to studying a possible evolutionary connection between a starburst and an active galactic nuclei (Mouri \& Taniguchi 2002 and references therein). The latter harbors a supermassive BH as the central engine.

Since the 2 \micron\ secondary peak of M82 has a starburst age of order $10^7\,{\rm yr}$ (Satyapal et al. 1997), the formation of the intermediate-mass BH should have occurred within this short duration. There is a possibility of runaway merging, which is known for collisions of BHs and those of normal stars (Quinlan \& Shapiro 1989; Lee 1993; Portegies Zwart et al. 1999). Massive objects segregate into the cluster core due to dynamical friction, increasing the merging probability. The probability itself is higher for more massive objects. Their merging products are even more massive. Thus the growth rate is higher than that expected from the merging rate at the initial time. 

However, runaway merging is a nonlinear process and hence has been studied only for specific cases using numerical methods, e.g., $N$-body simulation. Theoretical understanding of the underlying physics is insufficient. Also, the existing numerical results are not applicable to formation of an intermediate-mass BH. A more general analytical approach, albeit less realistic, is desirable.

We make an analytical analysis of the runaway merging. In our idealized model, BHs of a single mass $m_0$ exist initially with a number density $n_0$ and a velocity dispersion $v_0$. These BHs are assumed to merge with each other via two-body interactions, i.e., energy loss due to gravitational radiation (\S2). The BH mass distribution is studied using the Smoluchowski equation, a master equation describing the evolution of number densities of various-mass particles that merge with each other via two-body interactions. We obtain an analytical constraint on the timescale of runaway merging (\S3), and apply the result to BHs in a super star cluster with $m_0 = 30\,M_{\sun}$, $n_0 = 10^6\, {\rm pc}^{-3}$, and $v_0 = 1\, \mbox{km\,s}^{-1}$ (\S4).

\section{MERGING VIA GRAVITATIONAL RADIATION}

When two BHs become close, energy loss due to gravitational radiation could exceed the orbital kinetic energy, and a binary could form. The binary immediately merges through subsequent gravitational radiation. For this process, we estimate the cross section and then the rate coefficient.

We consider an interaction of two BHs where the BH masses are $m_i$ and $m_j$, the initial relative velocity is $v_{{\rm rel}}$, and the impact parameter is $b$. If the two BHs pass away narrowly, the orbit is close to parabolic. We thereby assume that the eccentricity $e$ is slightly greater than 1, and consider the leading term for $e \rightarrow 1$. This is equivalent to an assumption of significant gravitational focusing. The distance of closest approach $r_{{\rm min}}$ is
\begin{equation}
\label{eq1}
r_{{\rm min}} = \frac{b^2 v_{{\rm rel}}^2}{2 G (m_i+m_j)}.
\end{equation}
For an unperturbed parabolic ($e = 1$) or hyperbolic ($e > 1$) orbit, the quadrupole formalism gives the total energy of the gravitational radiation $\delta E$ at Newtonian order as
\begin{equation}
\label{eq2}
\delta E = \frac{8}{15}
           \frac{G^{7/2}}{c^5}
           \frac{(m_i+m_j)^{1/2} m_i^2 m_j^2}{r_{{\rm min}}^{7/2}}
           g(e),
\end{equation}
with $g(e) = 425\pi/(32\sqrt{2})$ at $e = 1$ (Turner 1977). The condition for the two BHs to pass away is
\begin{equation}
\label{eq3}
\delta E < \frac{1}{2} \frac{m_i m_j}{m_i+m_j} v_{{\rm rel}}^2.
\end{equation}
Equations (\ref{eq1})--(\ref{eq3}) yield the minimum impact parameter $b_{\rm min}$ for the BHs to pass away. Then we obtain the merging cross section $\sigma_{{\rm mer}} = \pi b_{{\rm min}}^2$ as
\begin{equation}
\label{eq4}
\sigma_{{\rm mer}} = 2 \pi \left( \frac{85\pi}{6\sqrt{2}} \right)^{2/7}
                       \frac{G^2 (m_i+m_j)^{10/7} m_i^{2/7} m_j^{2/7}}
                            {c^{10/7} v_{{\rm rel}}^{18/7}}. 
\end{equation}
Quinlan \& Shapiro (1989) derived the same formula with a somewhat different reasoning. The cross section $\sigma_{{\rm mer}}$ is larger than that corresponding to the Schwarzschild radius $\sigma_{{\rm Sch}} \simeq \pi (2Gm/c^2)^2 \simeq (v_{{\rm rel}}/c)^{18/7} \sigma_{{\rm mer}}$. Thus our Newtonian approximation is sufficiently accurate.

Since the merging cross section $\sigma_{{\rm mer}}$ is smaller than that for gravitational scattering, $\sigma_{{\rm sca}} \simeq \pi (Gm/v_{{\rm rel}}^2)^2 \simeq (c/v_{{\rm rel}})^{10/7} \sigma_{{\rm mer}}$, the BH motions are assumed to be in thermal equilibrium as a zeroth-order approximation. The equipartition of the kinetic energy is assumed to be achieved as
\begin{equation}
\label{eq5}
\frac{1}{2} m_i v_i^2 = \frac{1}{2} m_0 v_0^2 \quad \mbox{for any $m_i$}.
\end{equation}
Here $v_i$ and $v_0$ are three-dimensional velocity dispersions for BH masses $m_i$ and $m_0$, respectively. The probability distribution of the initial relative velocity $v_{{\rm rel}}$ is given by
\begin{equation}
\label{eq6}
P(x) = \left( \frac{x}{2\pi} \right) ^{1/2} \exp \left( -\frac{x}{2} \right) 
\quad \mbox{with} \quad
x = \frac{3 m_i m_j v_{{\rm rel}}^2}{(m_i+m_j)m_0 v_0^2}.
\end{equation}
Equations (\ref{eq4}) and (\ref{eq6}) yield the average $\langle v_{{\rm rel}}\, \sigma_{{\rm mer}} \rangle$ as the rate coefficient $R_{i,j}$ for merging of BHs of masses $m_i$ and $m_j$:
\begin{equation}
\label{eq7}
R_{i,j} = \frac{A G^2 m_0^2}{c^3}
              \left( \frac{v_0}{c}         \right) ^{-11/7}
              \left( \frac{m_i m_j}{m_0^2} \right) ^{15/14}
              \left( \frac{m_i+m_j}{m_0}   \right) ^{9/14},
\end{equation}
with $A = \Gamma(5/7)\,(2\pi)^{11/14}\,3^{1/2}\,85^{2/7} \simeq 33.33$. However, in this estimate, mass segregation has not been taken into account. Massive BHs sink into the core of the star cluster and have a high number density even if their total number is small. The scale height of BHs of a mass $m_i$ is proportional to $m_i^{-1/2}$ (multimass King's model; Sigurdsson \& Phinney 1995). Thus mass segregation enhances the merging rate by a factor $(m_i m_j/m_0^2)^{3/2}$. The rate coefficient is then proportional to $(m_i m_j/m_0^2)^{18/7}$ instead of $(m_i m_j/m_0^2)^{15/14}$.

The expression (\ref{eq7}) is too complicated to be studied analytically using the Smoluchowski equation. We use the inequality $m_i + m_j \ge 2 (m_i m_j)^{1/2}$ and obtain a simplified expression for the merging rate coefficient:
\begin{equation}
\label{eq8}
R_{i,j} \ge \frac{B G^2 m_0^2}{c^3}
                \left( \frac{v_0}{c}         \right) ^{-11/7}
                \left( \frac{m_i m_j}{m_0^2} \right) ^{\lambda/2},
\end{equation}
with $B = \Gamma(5/7)\,\pi^{11/14}\,2^{10/7}\,3^{1/2}\,85^{2/7} \simeq 52.04$. The exponent $\lambda$ is $39/14$ or $81/14$, respectively, if we ignore or incorporate the mass segregation.

We consider two-body interactions alone and thus ignore three-body interactions. They result in BH binaries. The BH binary undergoes subsequent interactions with other BHs and increases its binding energy. At each of the interactions, a fraction of the change in binding energy is converted into the translational kinetic energy. The BH binary eventually merges by emitting gravitational radiation or escapes from the star cluster (Sigurdsson \& Hernquist 1993; see also Portegies Zwart \& McMillan 2000). Although a numerical calculation based on a reliable BH mass function is required to draw a definitive conclusion, we expect that the former occurs much more often than the latter in a massive cluster with a large escape velocity if the BHs are massive enough ($\gg 10\,M_{\sun}$; Miller \& Hamilton 2001). The merging timescale is short for a binary of massive BHs. The escape of a massive BH binary is due to a collision with a massive BH. Before such a rare collision, the BH binary would merge. Our present model would accordingly underestimate the merging rate. This possible underestimation is not serious because we are to obtain an upper limit on the timescale of the runaway merging.

\section{SMOLUCHOWSKI EQUATION}

Suppose that particles of a single mass $m_0$ exist with a number density $n_0$ at the initial time $t=0$. They merge with each other via two-body interactions. The merging of particles of masses $m_i$ and $m_j$ has a rate coefficient $R_{i,j}$ and results in a particle of a mass $m_i + m_j$. We consequently have particles of masses $m_i = m_0 i$ ($i = 1,2,3,...$) and number densities $n_i$. If nondimensional terms are defined as
\begin{equation}
\label{eq9}
\tilde{n}_i     = \frac{n_i}    {n_0},\quad
\tilde{R}_{i,j} = \frac{R_{i,j}}{R_{0,0}},\quad \mbox{and} \quad
\tilde{t}       = n_0 R_{0,0}t,
\end{equation}
the evolution of the number densities is described by the Smoluchowski equation:
\begin{equation}
\label{eq10}
\frac{d \tilde{n}_i}{d \tilde{t}}  =
\frac{1}{2} \sum_{j+k=i} \tilde{R}_{j,k} \tilde{n}_j \tilde{n}_k
-\tilde{n}_i \sum_{j=1}^{\infty} \tilde{R}_{i,j} \tilde{n}_j.
\end{equation}
Here the first sum accounts for the increase of $\tilde{n}_i$ due to merging of particles satisfying the condition $m_j + m_k = m_i$ while the second sum accounts for the decrease of $\tilde{n}_i$ due to merging of particles of a mass $m_i$ with those of any mass. The initial condition is $\tilde{n}_1 = 1$ and $\tilde{n}_i = 0$ for $i \ge 2$ at $\tilde{t} = 0$.

The Smoluchowski equation is based on a mean-field approximation and thereby ignores spatial fluctuations, e.g., mass segregation. It is nevertheless possible to incorporate consequences of the fluctuation into the rate coefficient $R_{i,j}$. The number densities $n_i$ are then regarded as those averaged over the fluctuation scale for particles of a mass $m_0$.

The analytical solution of equation (\ref{eq10}) is known only for $\tilde{R}_{i,j} = 1$, $(i+j)/2$, and $ij$ (see Hayashi \& Nakagawa 1975). We study the solution for $\tilde{R}_{i,j} = (ij)^{\lambda/2}$ with $\lambda \ge 2$ and relate it to the solution for $\tilde{R}_{i,j} \ge (ij)^{\lambda/2}$, which corresponds to our rate coefficient (\ref{eq8}), using the $\ell$-th moment:
\begin{equation}
\label{eq11}
\tilde{M}_{\ell} = \sum_{i=1}^{\infty} i^{\ell} \tilde{n}_i.
\end{equation}
The total mass density and average mass of the particles are given respectively by $\tilde{M}_1$ and $\tilde{M}_2/\tilde{M}_1$ (Hayakawa \& Hayakawa 1988). From equation (\ref{eq10}), the moment $\tilde{M}_{\ell}$ evolves as 
\begin{equation}
\label{eq12}
\frac{d \tilde{M}_{\ell}}{d \tilde{t}} =
\frac{1}{2} \sum_{i,j} \tilde{n}_i 
                       \tilde{n}_j 
                       \tilde{R}_{i,j}
                       \left[ (i+j)^{\ell} - i^{\ell} -j^{\ell} \right].
\end{equation}
We have $d \tilde{M}_1 / d \tilde{t} = 0$, which means the conservation of the total mass, $\tilde{M}_1=1$. Multiplying both sides of equation (\ref{eq12}) for $\tilde{R}_{i,j} = (ij)^{\lambda/2}$  and $\ell = 2$ by $\tilde{M}_2^{-\lambda}$ gives
\begin{equation}
\label{eq13}
\frac{d \tilde{M}_2^{-(\lambda -1)}}{d \tilde{t}} \le -(\lambda -1)
\quad \mbox{for $\lambda \ge 2$}.
\end{equation}
Here we have used the mass conservation $\tilde{M}_1 = 1$ and also a relation originating in the Schwarz inequality:
\begin{equation}
\label{eq14}
\displaystyle
\frac{\sum_{i=1}^{\infty} i^{\lambda/2 +1} \tilde{n}_i}
     {\sum_{i=1}^{\infty}               i  \tilde{n}_i}
\ge
\left( \frac{\sum_{i=1}^{\infty} i^2 \tilde{n}_i}
            {\sum_{i=1}^{\infty}   i \tilde{n}_i} \right) ^{\lambda/2}
\quad \mbox{for $\lambda \ge 2$}.
\end{equation}
The equality holds at $\lambda = 2$. Equation (\ref{eq13}) has the solution:
\begin{equation}
\label{eq15}
\tilde{M}_2 \ge \left[ \frac{1}{1-(\lambda -1)\tilde{t}} \right]^{\frac{1}{\lambda -1}}
\quad \mbox{for $\lambda \ge 2$}.
\end{equation}
Thus the average mass $\tilde{M}_2/\tilde{M}_1 = \tilde{M}_2$ becomes infinity at a finite time $\tilde{t}_{\rm rm} \le (\lambda -1)^{-1}$, owing to emergence of infinite-mass particles. We regard $\tilde{t}_{\rm rm}$ as the timescale of runaway merging, which is shorter than the timescale $\tilde{t}=1$ corresponding to the initial merging rate $n_0 R_{0,0}$ (eq. [\ref{eq9}]). Even if $\tilde{R}_{i,j} = (ij)^{\lambda/2}$ is replaced with $\tilde{R}_{i,j} \ge (ij)^{\lambda/2}$, the solution (\ref{eq15}) remains the same.

The total number density of the particles is given by the zeroth moment $\tilde{M}_0$. Hence it would appear natural to adopt $\tilde{M}_1/\tilde{M}_0 = \tilde{M}_0^{-1}$ as the average mass. If both sides of equation (\ref{eq12}) for $\tilde{R}_{i,j} \ge (ij)^{\lambda/2}$ with $\lambda \ge 2$ and $\ell = 0$ are multiplied by $\tilde{M}_0^{\lambda -2}$, it follows that $\tilde{M}_0$ becomes zero at a finite time $\tilde{t}'_{{\rm rm}} \le 2(\lambda -1)^{-1}$. However, before the time $\tilde{t}'_{{\rm rm}}$, the moment equation $\tilde{M}_1 = 1$ becomes invalid (Hendriks, Ernst, \& Ziff 1983). This is the reason why we have adopted $\tilde{M}_2/\tilde{M}_1$ as the average mass.

The emergence of infinite-mass particles occurs in more general cases. If the rate coefficient $\tilde{R}_{i,j}$ scales as $\tilde{R}_{ai,aj} = a^{\lambda} \tilde{R}_{i,j} = a^{\lambda} \tilde{R}_{j,i}$, infinite-mass particles emerge within a finite timescale for $\lambda > 1$ (Hayashi \& Nakagawa 1975). Lee (1993) used this property to interpret runaway merging of BHs found in $N$-body simulations. Since the timescale of the runaway merging had not been estimated analytically, we have obtained the analytical constraint (see Hendriks et al. 1983 for similar discussion in terms of sol-gel phase transition). If the rate coefficient behaves as $\tilde{R}_{i,j} \sim i^{\mu} j^{\nu}$ with $\nu > 1$ for $j \rightarrow \infty$, infinite-mass particles emerge and high-order moments become infinity just after the initial time $\tilde{t} = 0$ (van Dongen 1987). This is the case in our rate coefficient (\ref{eq8}). We have ignored moments $\tilde{M}_{\ell}$ with $\ell > 2$ because the Smoluchowski equation (\ref{eq10}) is merely used as an idealized model in the present work.

\section{Discussion}

The runaway merging of BHs is caused by two mechanisms, which are equally important. First, the merging rate $R_{i,j}$ is intrinsically large for massive BHs (eq. [\ref{eq7}]). This is because the energy loss due to gravitational radiation $\delta E$ is large. Massive BHs have low velocities (eq. [\ref{eq5}]). If the initial relative velocity $v_{{\rm rel}}$ is low or the BH masses $m_i$ and $m_j$ are large, the gravitational focusing is significant and the distance of closest approach $r_{{\rm min}}$ is small (eq. [\ref{eq1}]). If the distance $r_{{\rm min}}$ is small or the BH masses $m_i$ and $m_j$ are large, the energy loss $\delta E$ is large (eq. [\ref{eq2}]). Second, low-velocity massive BHs sink into the cluster core and have an enhanced number density. The first and second mechanisms correspond to $\lambda = 39/14$ and 3, respectively. They are comparable and individually exceed the critical value for the onset of runaway merging, $\lambda = 1$ (\S3).

Equations (\ref{eq8}), (\ref{eq9}), and (\ref{eq15}) yield a constraint on the timescale $t_{\rm rm}$ of the runaway merging as
\begin{eqnarray}
\label{eq16}
t_{\rm rm} & \le & \frac{1}{\lambda -1} 
                  \frac{c^3}{B G^2 m_0^2 n_0}
                  \left( \frac{v_0}{c} \right) ^{11/7} \\
          &  =  & \left( \frac{m_0}{30\, M_{\sun}} \right) ^{-2}
                  \left( \frac{n_0}{10^6\, {\rm pc}^{-3}} \right) ^{-1}
                  \left( \frac{v_0}{1\, \mbox{km\,s}^{-1}} \right) ^{11/7}
                  \times \left\{ \begin{array}{ll}
                   4 \times 10^7\ {\rm yr} \quad \mbox{no mass segregation} \\
                   2 \times 10^7\ {\rm yr} \quad \mbox{mass segregation.}
                  \end{array} \right. \nonumber
\end{eqnarray}
The mass $m_0 = 30\,M_{\sun}$, the number density $n_0 = 10^6\, {\rm pc}^{-3}$, and the velocity dispersion $v_0 = 1\, \mbox{km\,s}^{-1}$ provide the most favorable conditions for the runaway merging among those expected for BHs in actual super star clusters (see below). We accordingly obtain an upper limit of order $10^7\,{\rm yr}$ on the timescale $t_{\rm rm}$. This upper limit is comparable to the starburst age observed for the 2 \micron\ secondary peak of the starburst galaxy M82 (\S1). Since a starburst duration is of order $10^8\,{\rm yr}$ (see de Grijs, O'Connell, \& Gallagher 2001 for the case of a fossil starburst M82 B), it is generally possible that an intermediate-mass BH is formed in an ongoing starburst region.

The BH mass $m_0=30\,M_{\sun}$ has been adopted because stars heavier than 30--$40\,M_{\sun}$ evolve directly to BHs with no supernova explosion (Brown \& Bethe 1994; Fryer 1999). The BH mass is equal to that of the progenitor. Although the less massive stars could evolve to BHs, they explode as supernovae and blow off most of their mass. A low-mass BH is easily ejected from the star cluster by three-body interactions and does not contribute to the intermediate-mass BH (\S2). 

The number density $n_0 = 10^6\, {\rm pc}^{-3}$ and the velocity dispersion $v_0 = 1\, \mbox{km\,s}^{-1}$ of the BHs have been adopted in the light of a nearby super star cluster, R136 in the Large Magellanic Cloud (see Portegies Zwart et al. 1999). The central mass density of R136 is $10^6\,M_{\sun}\,{\rm pc}^{-3}$. We assume the Salpeter initial mass function $dn/dm \propto m^{-2.35}$ with the cutoffs at 0.1 and $100\,M_{\sun}$, which yields the number fraction 0.04\% for stars heavier than $30\,M_{\sun}$ and the average mass $0.4\,M_{\sun}$. We also assume the multimass King's model (\S2). Then the above mass density corresponds to the number density $n_0 \gtrsim 10^6\,{\rm pc}^{-3}$ for BHs heavier than $30\,M_{\sun}$. Here wind mass loss from stars has been ignored as a zeroth-order approximation. The mass loss rate observed for a $30\,M_{\sun}$ star is $10^{-6}\,M_{\sun}\,{\rm yr}^{-1}$ (Becker 1999). Since the lifetime of a $30\,M_{\sun}$ star is $10^6$ yr, the total mass loss is at most a few times $M_{\sun}$. On the other hand, the stellar velocity dispersion of R136 is estimated to be about $10\,\mbox{km\,s}^{-1}$, which corresponds to the velocity dispersion $v_0 \simeq 1\, \mbox{km\,s}^{-1}$ for BHs of a mass $30\,M_{\sun}$. If runaway merging occurs among stars before they evolve to BHs (\S1), the number density $n_0$ could be higher and hence the upper limit on the timescale $t_{\rm rm}$ could be lower.

The timescale $t_{{\rm ms}}$ of mass segregation is less than that of the runaway merging. If the average stellar mass $m$ is $0.4\,M_{\sun}$, the mass density $\rho$ is $10^6\,M_{\sun}\,{\rm pc}^{-3}$, and the velocity dispersion $v$ is $10\,\mbox{km\,s}^{-1}$, the star cluster has the relaxation timescale $t_{{\rm rlx}} \simeq v^3/(\pi G^2 m \rho) = 4 \times 10^7\,{\rm yr}$, which yields $t_{{\rm ms}} \simeq (m/m_0) t_{{\rm rlx}} = 5 \times 10^5\,{\rm yr}$ for BHs of a mass $m_0 = 30\,M_{\sun}$ (Spitzer 1969). 

Super star clusters have ranges of structure parameters. Existing few observations of super star clusters other than R136 tend to imply $v_0 > 1\, \mbox{km\,s}^{-1}$ (Ho \& Filippenko 1996; Smith \& Gallagher 2001). Observations of globular clusters, i.e., likely descendants of super star clusters, tend to imply $n_0 < 10^6\,{\rm pc}^{-3}$ (Pryor \& Meylan 1993). In these cases, the timescale $t_{{\rm rm}}$ of the runaway merging is long (eq. [\ref{eq16}]). The formation of an intermediate-mass BH in an ongoing starburst region occurs only under rare favorable conditions, i.e., high number density $n_0$ and low velocity dispersion $v_0$ of stellar-mass BHs. Also, the star cluster as a whole has to be sufficiently massive in order to provide a sufficient number of the stellar-mass BHs. In fact, only one intermediate-mass BH heavier than $10^3\,M_{\sun}$ has been found among more than 100 super star clusters of the starburst galaxy M82, although there might exist some intermediate-mass BHs that are not accreting the gas and hence are not observable.

%\clearpage

\end{document}